\documentclass[12pt,a4paper,dvips,final]{article}
\usepackage{epsfig,times}  

\renewcommand{\section}[1]{\noindent\mbox{\bf{#1}} \newline\noindent}
\newcommand{\secref}[1]{\noindent\mbox{\bf{#1}}}
\renewcommand{\subsection}[1]{\noindent\mbox{\it{#1}} \newline\noindent}
\renewcommand{\subsubsection}[1]{\noindent\mbox{#1}--}

\newfont{\tf}{cmbx10 at 16pt}
\newfont{\af}{cmr10 at 14pt}

\def\fwb{70mm}
\def\fhb{65mm}

\newcommand{\BC}{$B/C$}
\newcommand{\Berat}{$^{10}Be/\,^9Be$}
\newcommand{\Dxx}{D_{xx}}

\begin{document}

\frenchspacing

{\tf \noindent New constraints on Galactic cosmic-ray propagation \\}

{\af \noindent A.W. Strong$^1$ and I.V. Moskalenko$^{1,2}$ \\}

{\it \noindent
$^1$Max-Planck-Institut f\"ur extraterrestrische Physik, 
D-85740 Garching, Germany \\
$^2$Institute for Nuclear Physics, Moscow State University, 
119 899 Moscow, Russia }\\

\section{Abstract}
An extensive program for the calculation of galactic cosmic-ray
propagation has been developed.  Primary and secondary nucleons,
primary and secondary electrons, secondary positrons and antiprotons
are included.  Fragmentation and energy losses are computed using
realistic distributions for the interstellar gas and radiation fields.

Models with diffusion and convection only do not account naturally for
the observed energy dependence of \BC, while models with reacceleration
reproduce this easily. The height of the halo propagation region is
determined, using recent \Berat\ measurements, as greater than 4 kpc.
The radial distribution of cosmic-ray sources required is broader than
current estimates of the SNR distribution for all halo sizes.  Our
results include an estimate of cosmic-ray antiproton and positron
spectra, and the Galactic diffuse $\gamma$-ray emission (see
accompanying paper: Moskalenko 1998b).
\\

\section{Introduction.}
We are constructing a model which aims to reproduce self-consistently
observational data of many kinds related to cosmic-ray (CR) origin and
propagation: direct measurements of nuclei, electrons, positrons,
antiprotons, gamma rays, and synchrotron radiation. These data provide
many independent constraints on any model and our approach is able to
take advantage of this since it must be consistent with all types of
observation.

Here we present our results on the evaluation of diffusion/convection
and reacceleration models based on the \BC\ and \Berat\ ratios, and set
limits on the halo size.  A re-evaluation of the halo size is desirable
since new \Berat\ data are now available from Ulysses (Connell 1998)
with better statistics than previously.  Our preliminary results were
presented in Strong (1997a,b) and full results for protons, Helium,
positrons, and electrons in Moskalenko (1998a).  Some illustrative
results for gamma-rays and synchrotron radiation are given in Strong
(1997a) and Moskalenko (1998b) and all details are given in
Strong (1998)\footnote{For more details see
{\it http://www.gamma.mpe--garching.mpg.de/$\sim$aws/aws.html}}.
\\

\section{The model description.}
The models are three dimensional with cylindrical symmetry in the
Galaxy, and the basic coordinates are $(R,z,p)$, where $R$ is
Galactocentric radius, $z$ is the distance from the Galactic plane, and
$p$ is the particle momentum. The propagation equations are solved
numerically on a grid by the method described in Strong (1998).
$R_\odot$ is taken as 8.5 kpc.  The propagation region is bounded by
$R=R_h$, $z=z_h$ beyond which free escape is assumed. We take $R_h=30$
kpc. The range $z_h=1-20$ kpc is considered. For a given $z_h$ the
diffusion coefficient as a function of momentum is determined by
\BC\ for the case of no reacceleration; if reacceleration is assumed
then the reacceleration strength (related to the Alfv\'en speed) is
constrained by the energy-dependence of \BC.  The spatial diffusion
coefficient for the case of no reacceleration is taken as $\Dxx = \beta
D_0(\rho/\rho_0)^{\delta_1}$ below rigidity $\rho_0$, $\beta
D_0(\rho/\rho_0)^{\delta_2}$ above rigidity $\rho_0$.  Since the
introduction of a sharp break in $\Dxx$ is an extremely contrived
procedure which is adopted just to fit \BC\ at all energies, we also
consider the case $\delta_1=\delta_2$ (no break).  The convection
velocity $V=V(z)$ is assumed to increase linearly with distance from
the plane.  For the case with reacceleration, the spatial diffusion
coefficient is $\Dxx = \beta D_0(\rho/\rho_0)^\delta$ with
$\delta=\frac{1}{3}$ for all rigidities, and the momentum-space
diffusion coefficient $D_{pp}$ is related to $\Dxx$ (using Berezinskii
1990 and Seo 1994).  The injection spectrum of nucleons is assumed to
be a power law in momentum.  The interstellar hydrogen distribution
uses HI and CO surveys and information on the ionized component; the
Helium fraction of the gas is taken as 0.11 by number.  Energy losses
for nucleons by ionization and Coulomb interactions are included.  The
distribution of CR sources is chosen to reproduce the
CR distribution determined by analysis of EGRET $\gamma$-ray
data (Strong 1996b).  The secondary nucleon and secondary $e^\pm$
source functions are computed from the propagated primary distribution
and the gas distribution.
\\

\begin{figure}[t!]
   \begin{picture}(148,73)(0,1)
      \put(0,0){\makebox(70,0)[lb]
         {\psfig{file=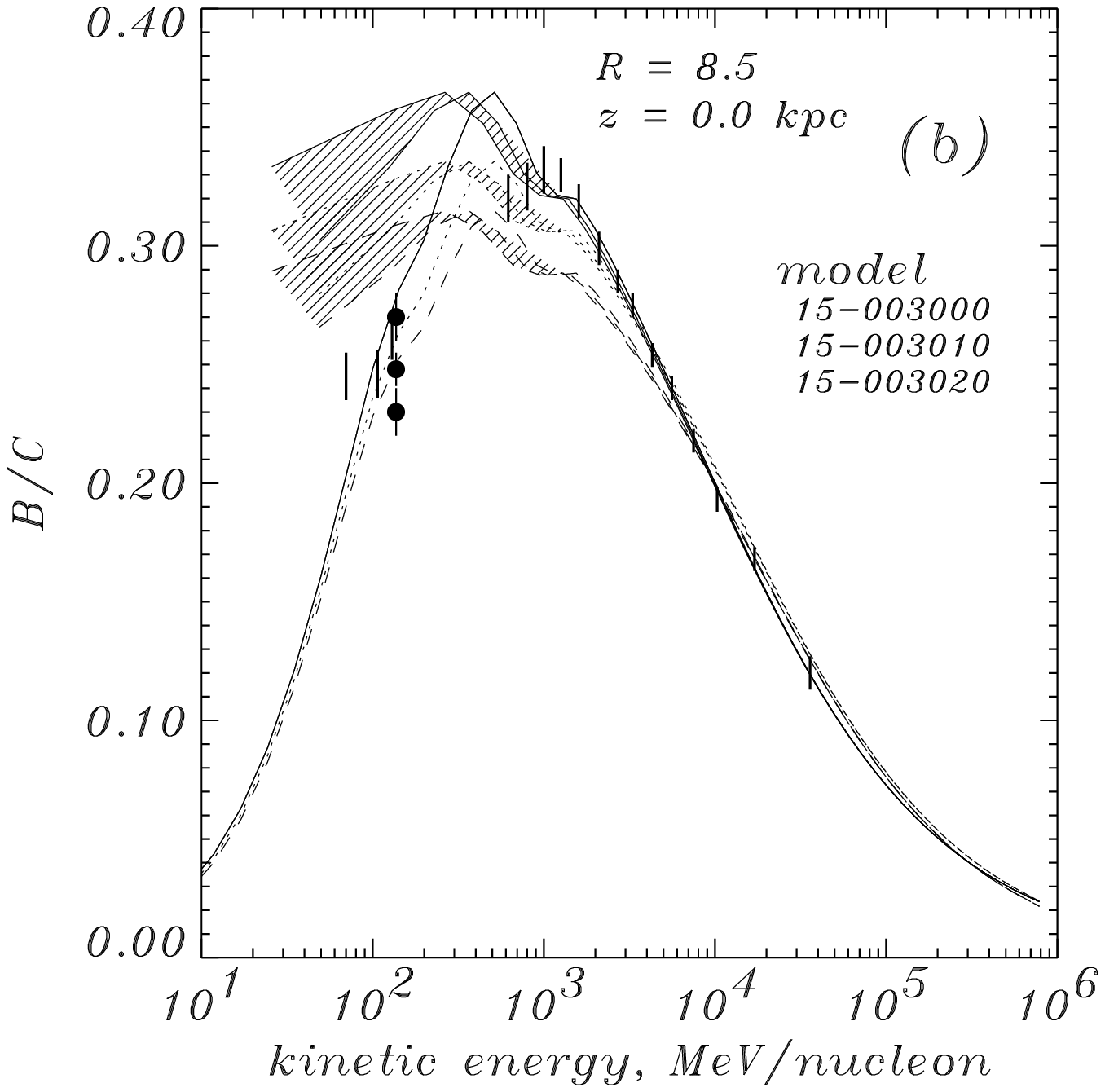,width=\fwb,height=\fhb,clip=}}}
      \put(75,0){\makebox(70,0)[lb]
         {\psfig{file=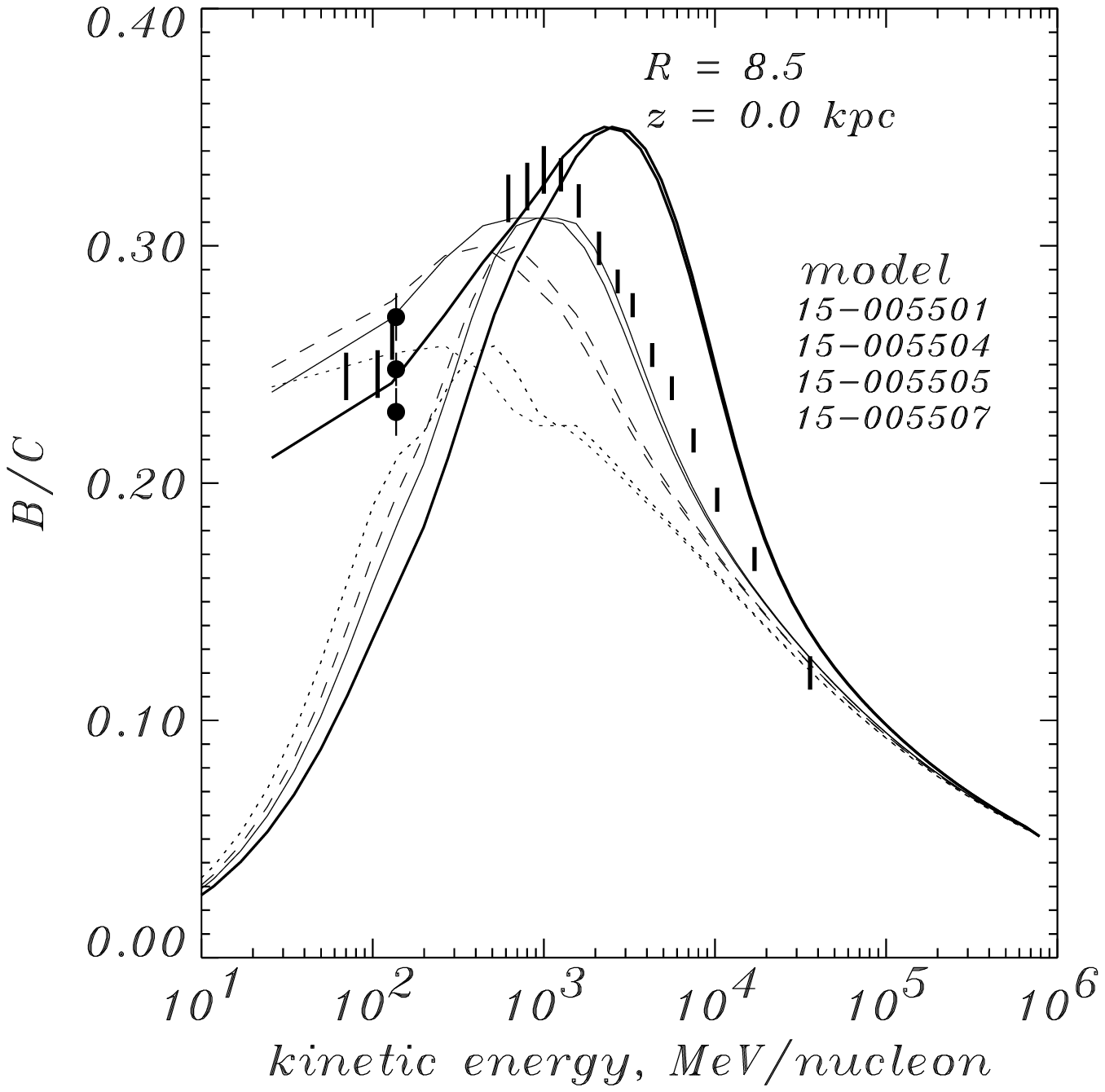,width=\fwb,height=\fhb,clip=}}}
   \end{picture}
\small Fig. 1.
{\it Left panel:} \BC\ ratio for diffusion/convection models without
break in diffusion coefficient (Strong 1998 and references therein),
for $z_h$ = 3 kpc, $dV/dz$ = 0 (solid line), 5 (dotted line), and 10 km
s$^{-1}$ kpc$^{-1}$ (dashed line); solid line: interstellar ratio,
shaded area: modulated to 300 -- 500 MV; data: HEAO-3, Voyager,
Ulysses.
{\it Right panel:} \BC\ ratio for diffusive reacceleration models
(Strong 1998) with $z_h$ = 5 kpc, $v_A$ = 0 (dotted), 15 (dashed), 20
(thin solid), 30 km s$^{-1}$ (thick solid).  In each case the
interstellar ratio and the ratio modulated to 500 MV is shown.
\end{figure} 

\section{Illustrative results.}
We consider the cases of diffusion+convection and
diffusion+reacceleration, since these are the minimum combinations
which can reproduce the key observations. Our basic conclusion is that
the reacceleration models are more satisfactory in meeting the
constraints provided by the data, reproducing the \BC\ energy
dependence without {\it ad hoc} variations in the diffusion
coefficient; further it is not possible to find any {\it simple}
version of the diffusion/convection model which reproduces
\BC\ satisfactorily.

Figure~1a shows the diffusion+convection model without break,
$\delta_1 = \delta_2$; for each $dV/dz$, the remaining parameters
$D_0$, $\delta_1$ and $\rho_0$ are adjusted to fit the data as well as
possible.  It is clear that a {\it good} fit is {\it not} possible; the
basic effect of convection is to reduce the variation of \BC\ with
energy, and although this improves the fit at low energies the
characteristic peaked shape of the measured \BC\ cannot be reproduced.
If we allow $\delta_1\not=\delta_2$ it can clearly be fitted,
but the break has to be large and the procedure is {\it ad hoc}.

Figure~1b illustrates a diffusive reacceleration model and
shows the effect on \BC\ of varying $v_A$ ($= 0\div30$ km s$^{-1}$) for
$z_h= 5$ kpc.  This shows how the initial form becomes modified to
produce the characteristic peaked shape.  Reacceleration models thus
lead naturally to the observed peaked form of \BC, as pointed out by
previous authors (e.g., Letaw 1993, Seo 1994, Heinbach 1995);
a value $v_A\sim20$ km s$^{-1}$ seems satisfactory.

\begin{figure}[t!]
   \begin{picture}(148,73)(0,1)
      \put(0,0){\makebox(70,0)[lb]
         {\psfig{file=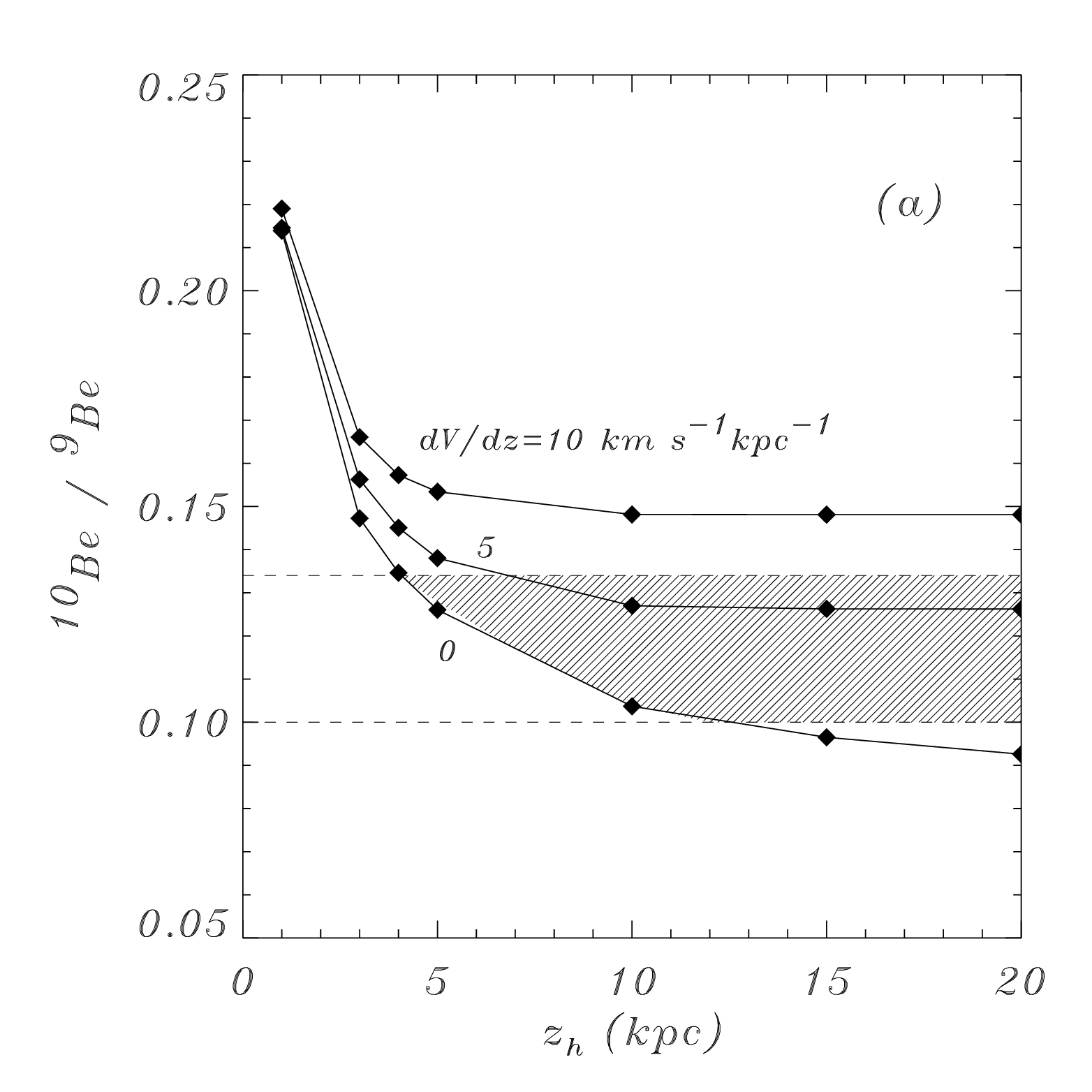,width=\fwb,height=\fhb,clip=}}}
      \put(75,0){\makebox(70,0)[lb]
         {\psfig{file=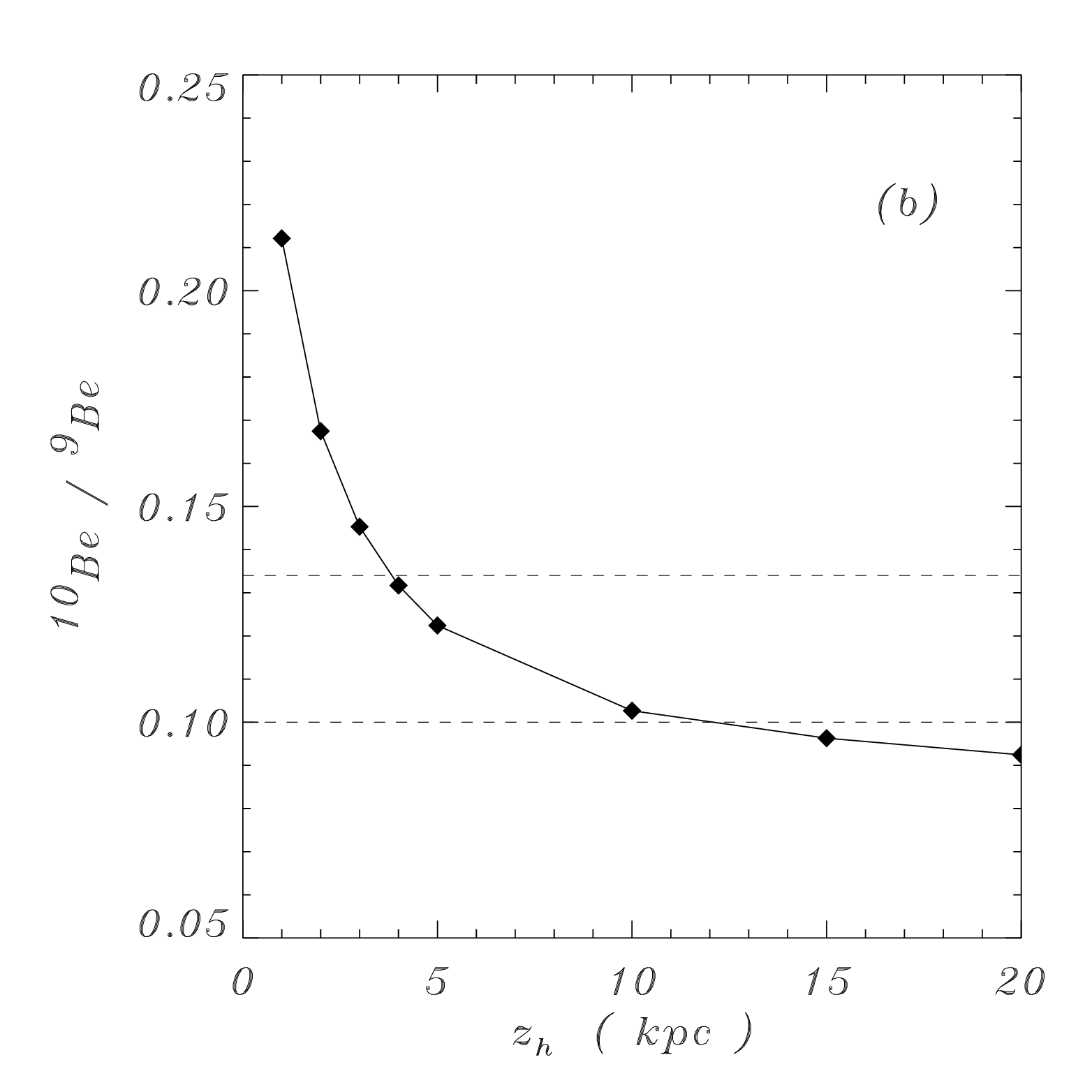,width=\fwb,height=\fhb,clip=}}}
   \end{picture}
\small Fig. 2. 
(a) Predicted \Berat\ ratio as function of $z_h$ for $dV/dz$ = 0, 5, 10
km s$^{-1}$ kpc$^{-1}$, the Ulysses experimental limits are shown as
horizontal dashed lines.  The shaded regions show the parameter ranges
allowed by the data.  (b) \Berat\ ratio for diffusive reacceleration
models as function of $z_h$ at 525 MeV/nucleon corresponding to the
mean interstellar value for the Ulysses data (Connell 1998).
\end{figure}

Figure~2a summarizes the limits on $z_h$ and $dV/dz$ for
diffusion/convection, using the \Berat\ ratio at the interstellar
energy of 525 MeV/nucleon appropriate to the Ulysses data (Connell
1998).  We conclude that in the absence of convection $4{\rm\ kpc}<z_h
< 12 {\rm\ kpc}$, and if convection is allowed the lower limit remains
but no upper limit can be set.  In the case $dV/dz < 7$ km s$^{-1}$
kpc$^{-1}$, this figure places upper limits on the convection parameter
for each halo size.  These limits are rather strict, and a finite wind
velocity is only allowed in any case for $z_h > 4$ kpc.  Figure~2b
shows \Berat\ for the reacceleration models as a function of $z_h$ at
525 MeV/nucleon corresponding to the Ulysses measurement and we
again find that $4{\rm\ kpc} <z_h < 12$ kpc.

Figure~3 (left panel) shows the effect of halo size on the
radial distribution of 3 GeV CR protons, for the reacceleration
model.  For comparison we show the CR distribution deduced by
model-fitting to EGRET gamma-ray data ($>100$ MeV) from Strong
(1996b), which is dominated by the $\pi^0$-decay component; the
analysis by Hunter (1997), based on a different approach, gives
a similar result.  The predicted CR distribution using the SNR
source function is too steep even for large halo sizes; in fact the
halo size has a relatively small effect on the distribution.  Other
related distributions such as pulsars have an even steeper falloff.
Based on these results we have to conclude, in the context of the
present models, that the distribution of sources is not that expected
from the (highly uncertain) distribution of SNR.  In view of the
difficulty of deriving the SNR distribution this is perhaps not a
serious shortcoming; if SNR are indeed CR sources then it is possible
that the gamma-ray analysis gives the best estimate of their Galactic
distribution. Therefore, we have chosen a CR source distribution
to fit the $\gamma$-ray data after propagation (Figure~3, right panel).
The possibility of anisotropic diffusion
(preferentially in the radial direction) has not yet been addressed in
our models.

\begin{figure}[t!]
   \begin{picture}(148,73)(0,1)
      \put(0,0){\makebox(70,0)[lb]
         {\psfig{file=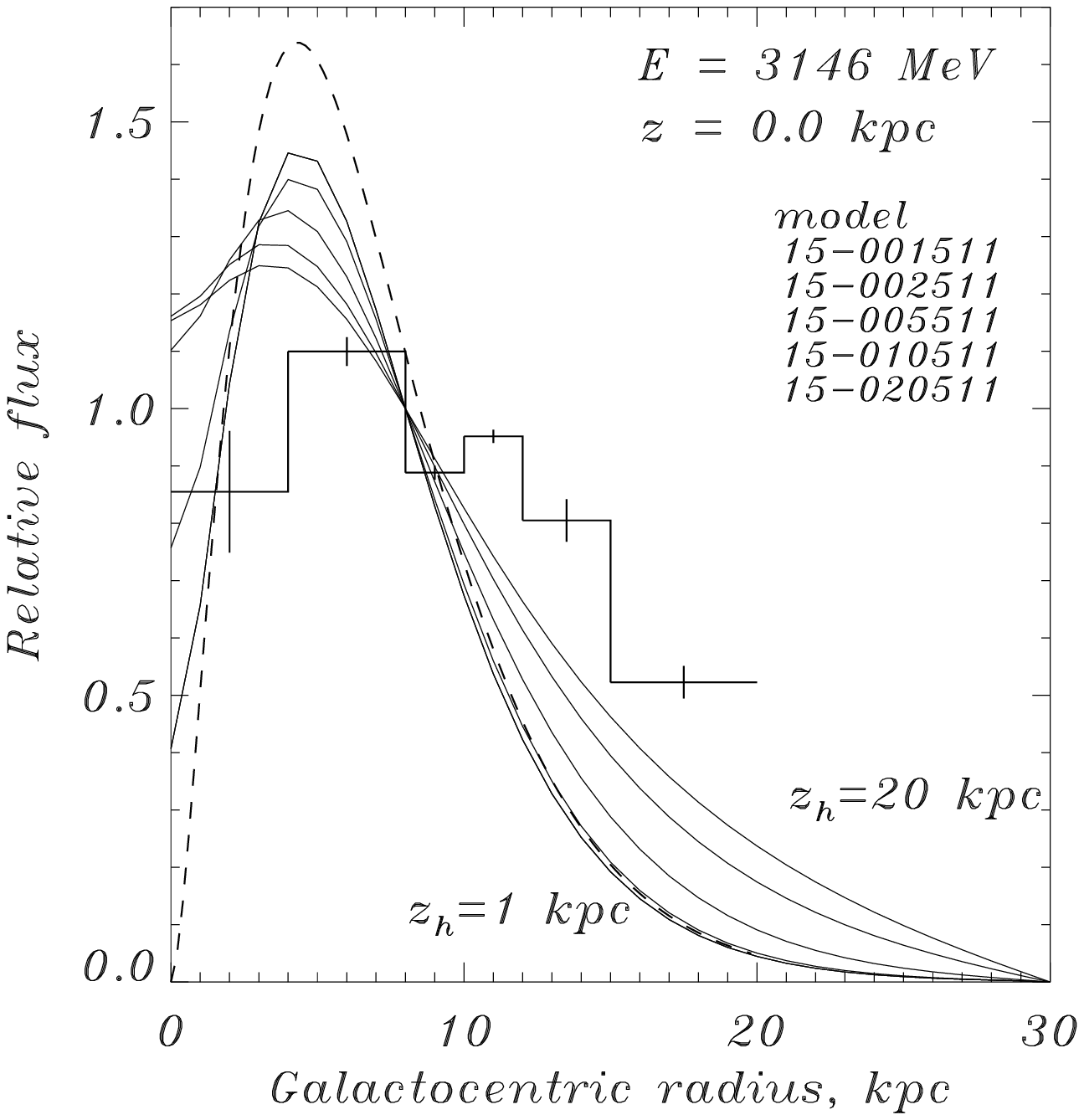,width=\fwb,height=\fhb,clip=}}}
      \put(75,0){\makebox(70,0)[lb]
         {\psfig{file=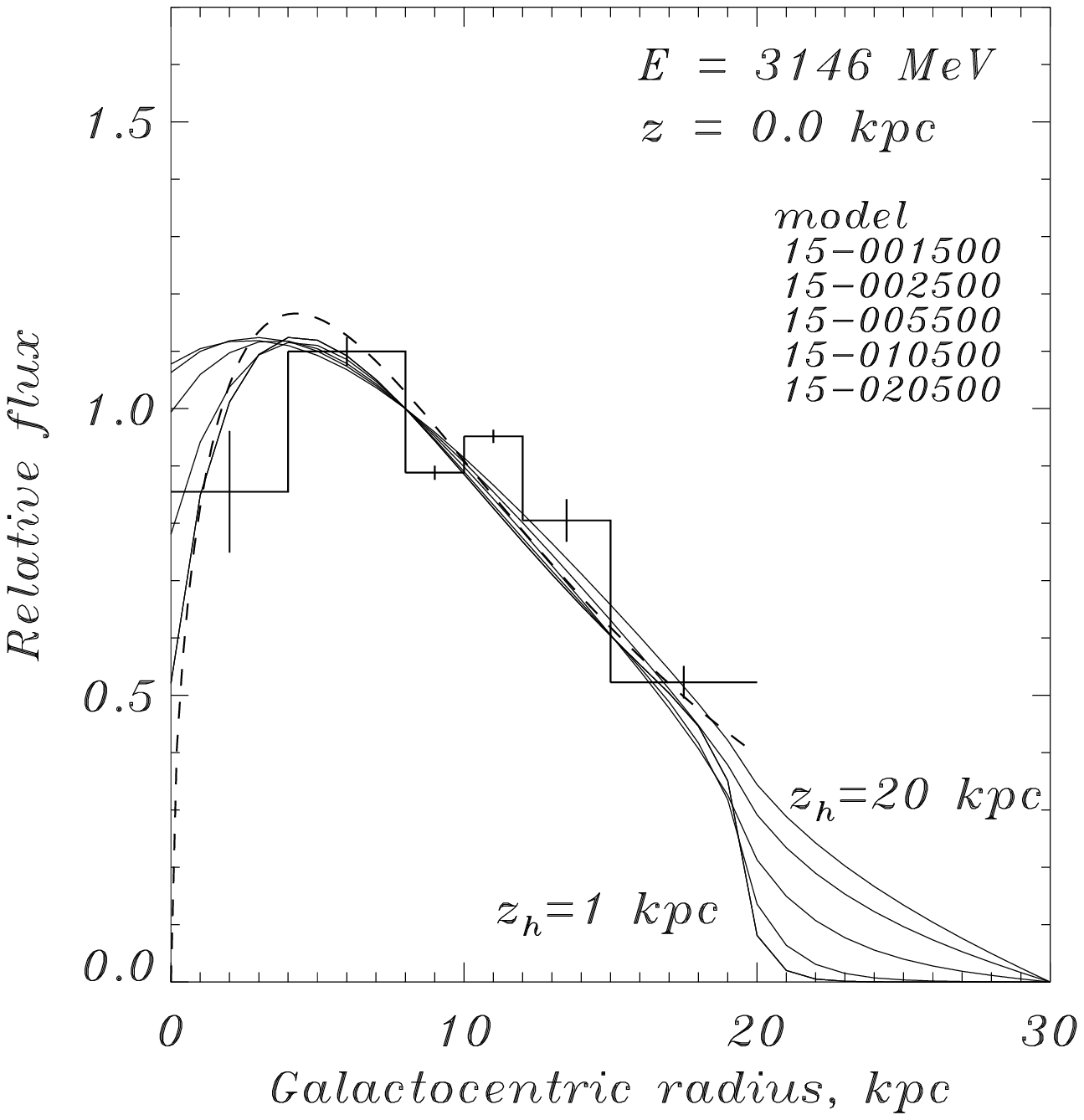,width=\fwb,height=\fhb,clip=}}}
   \end{picture}
\small Fig. 3. 
{\it Left panel}: radial distribution of 3 GeV protons at $z = 0$, for
diffusive reacceleration model with halo sizes $z_h = 1$, 3, 5, 10, 15,
and 20 kpc (solid curves). Dashed line: the source distribution is that
for SNR given by Case (1996),  histogram: the CR distribution
deduced from EGRET $>$100 MeV gamma rays (Strong 1996b).
{\it Right panel}:  radial distribution of 3 GeV protons at $z = 0$
for the source distribution actually adopted (dashed line),
for diffusive reacceleration model with various halo sizes $z_h = 1$,
3, 5, 10, 15, and 20 kpc (solid curves).
\label{fig4} 
\end{figure}

The positron fraction computed is in good agreement with the measured
one between 1 and 10 GeV, where the data are rather precise.  Our
positron predictions from Moskalenko (1998a) have been compared with
more recent absolute measurements in Barwick (1998) and the agreement
is good; for the positrons this new comparison has the advantage of
being independent of the electron spectrum (see also Moskalenko 1998b).
\\

\secref{References.} 
\setlength{\parindent}{-5mm}
\begin{list}{}{\topsep 0pt \partopsep 0pt \itemsep 0pt \leftmargin 5mm
\parsep 0pt \itemindent -5mm}
\item S.W.~Barwick et al.\ {\it Ap.J.} 498 (1998) 779--789.
\item V.S.~Berezinskii et al.\ {\it Astrophysics of Cosmic Rays.}
      North Holland.\ Amsterdam.\ (1990).
\item G.~Case and D.~Bhattacharya.\ {\it A\&AS.} 120C (1996) 437--440.
\item J.J.~Connell.\ {\it Ap.J.} 501 (1998) L59--L62.
\item U.~Heinbach and M.~Simon.\ {\it Ap.J.} 441 (1995) 209--221.
\item S.D.~Hunter et al.\ {\it Ap.J.} 481 (1997) 205--240.
\item J.R.~Letaw, R.~Silberberg and C.H.~Tsao.\ {\it Ap.J.} 414 (1993) 601--611.
\item I.V.~Moskalenko and A.W.~Strong.\ {\it Ap.J.} 493 (1998a) 694--707.
\item I.V.~Moskalenko and A.W.~Strong.\ {\it 16th ECRS.} (1998b) GR-1.3.\
   (astro-ph/9807288)
\item E.S.~Seo and V.S.~Ptuskin.\ {\it Ap.J.} 431 (1994) 705--714.
\item A.W.~Strong and J.R.~Mattox.\ {\it A\&A.} 308 (1996b) L21--L24.
\item A.W.~Strong and I.V.~Moskalenko.\ {\it 4th Compton Symp. AIP 410.} 
   Ed.\ C.D.~Dermer et al.\ 1162--1166.\ AIP.\ NY.\ (1997a).
\item A.W.~Strong, I.V.~Moskalenko and V.~Sch\"onfelder.\ 
      {\it 25th ICRC.} 4 (1997b) 257--260.
\item A.W.~Strong and I.V.~Moskalenko.\ {\it Ap.J.} 509 (1998) in press. 
      (astro-ph/9807150)

\end{list}
\end{document}